\documentclass[aps,prd,twocolumn,superscriptaddress,showpacs]{revtex4}

\usepackage{graphicx}

\begin{document}

\title{Comment on ``Heavy element production in inhomogeneous big bang
nucleosynthesis''}

\author{Thomas Rauscher}

\affiliation{Departement f\"ur Physik und Astronomie, Universit\"at Basel, 
CH-4056 Basel, Switzerland}


\date{\today}

\begin{abstract}
The work of Matsuura et al.\ [Phys.\ Rev.\ D {\bf 72}, 123505 (2005)]
claims that heavy nuclei could have been produced in a combined p- and
r-process in very high baryon density regions of an inhomogeneous big bang.
However, they do not account for observational constraints and previous
studies which show that such high baryon density regions did not
significantly contribute to big bang abundances.
\end{abstract}


\maketitle

The recent paper by Matsuura, Fujimoto, Nishimura, Hashimoto, and Sato
\cite{mfnhs} (hereafter referred to as MFNHS) presents reaction network
calculations of big bang nucleosynthesis (BBN) for baryon-to-photon ratios
$\eta=n_\mathrm{b}/n_\gamma \leq 10^{-2}$, far exceeding the commonly
adopted value of $\eta \approx 6\times 10^{-10}$. They argue that their study
can be motivated by supersymmetric models which may allow the creation of
bubbles with very high baryon density, approaching $\eta=1$. They find that
heavy elements, including p-nuclei, are synthesized in such scenarios with
high $\eta$.

In their studies, MFNHS neglect baryon diffusion. They motivate this by
stating that their aim is to study BBN in the
high-density regions created by supersymmetric baryogenesis and not to
make a precise adjustment between BBN and the measured Cosmic Microwave
Background Radiation (CMBR). Furthermore, they state that they just assume
that the high-density bubbles are large enough to neglect diffusion but not
so large as to contradict CMBR observations.

In MFNHS it is claimed that in order to compare the results with observations,
model-dependent dynamical mixing has to be invoked. However, it is possible
to make a first check of the feasibility of the BBN model by a simple
back-of-the-envelope calculation.
Even without considering details in the shape of the bubbles, a simple
estimate on the required properties of the bubbles can easily be made by
using the volume fraction $f_v$ occupied by the high-density bubbles and the
density contrast $R=n_\mathrm{b}^\mathrm{hi}/n_\mathrm{b}^\mathrm{lo}$
between regions of high and low density. The volume fraction $f_v$ can only
assume values in the range $0\leq f_v \leq 1$. When computing the BBN yields, one
has to average both density regions, leading to
\begin{equation}
\label{eq:mix}
\overline{X}_i \propto f_v R X_i^\mathrm{hi} + (1-f_v) X_i^\mathrm{lo} \quad ,
\end{equation}
with $X_i$ being the produced mass fraction of nucleus $i$.
This approach was already
introduced and used in \cite{ahs,fma,rau}.

Assuming a given value of the Hubble parameter
the $\eta$ values can be translated into a ratio of the baryon density
$\rho_\mathrm{b}$ to the critical density $\rho_\mathrm{c}$,
$\Omega_\mathrm{b}=\rho_\mathrm{b}/\rho_\mathrm{c}$.
The following
relations are obtained immediately:
\begin{eqnarray}
\Omega_\mathrm{b}=\overline{\Omega}_\mathrm{b}&=&f_v
\Omega_\mathrm{b}^\mathrm{hi} + (1-f_v) \Omega_\mathrm{b}^\mathrm{lo}
\quad,\\
\Omega_\mathrm{b}^\mathrm{hi}&=&\frac{R \Omega_\mathrm{b}}{f_v (R-1) + 1}
\quad,\\
\Omega_\mathrm{b}^\mathrm{lo}&=&\frac{\Omega_\mathrm{b}^\mathrm{hi}}{R} \quad .
\end{eqnarray}
The dependence of the density contrast $R$ on the volume fractions $f_v$
\begin{equation}
\label{eq:r}
R(f_v)=\frac{\Omega_\mathrm{b}^\mathrm{hi} (1-f_v)}{\Omega_\mathrm{b} -
\Omega_\mathrm{b}^\mathrm{hi} f_v}
\end{equation}
can easily be derived from the above, yielding positive values of $R$
only for
\begin{equation}
\Omega_\mathrm{b}^\mathrm{hi} f_v < \Omega_\mathrm{b} \quad .
\end{equation}

Table \ref{tab:omega} shows
$\Omega_\mathrm{b}^\mathrm{hi}$, the upper limit of the volume
fraction $f_v^\mathrm{max}$, and the lower limit of the density contrast
$R^\mathrm{min}$ for several values of $\eta$ used in MFNHS. The value of
$\Omega_\mathrm{b}^\mathrm{hi}$ was computed assuming that $\eta=6\times
10^{-10}$ corresponds to the standard BBN value of
$\Omega_\mathrm{b}\approx 0.05$. It is immediately obvious that the high
density regions can only occupy a tiny fraction
$0 < f_v < f_v^\mathrm{max}$ of the available space
when requiring the average $\Omega_\mathrm{b}$ to remain close to the
standard BBN value. From Eq.\ \ref{eq:r} it can be seen that the
corresponding density ratio range is
$R^\mathrm{min}=\Omega_\mathrm{b}^\mathrm{hi}/\Omega_\mathrm{b}\approx
20 \Omega_\mathrm{b}^\mathrm{hi} < R < \infty$. It has to be noted that
although the values shown in Tab.\ \ref{tab:omega} were derived assuming the
standard value of $\Omega_\mathrm{b}=0.05$, the numbers will not change
significantly even when, e.g., using the permitted maximal value
$\Omega_\mathrm{b}=0.3$ (this would imply, of course, that there is no dark matter
component in the Universe).

Two conclusions can be drawn from this simple estimate. First, the fact
that the volume fraction of the high density regions is so tiny renders
the assumption untenable that diffusion effects can be neglected.
Secondly, it becomes doubtful whether the observed light element
abundances can be reproduced in such a model. Since no values for the light
element abundances are given in MFNHS, one has to resort to previous works.
Fuller, Mathews, and
Alcock \cite{fma} have studied the BBN of light elements in detail as a
function of $f_v$ and $R$ in proton- as well as neutron-rich zones.
They find best agreement with primordial light element abundances for
$f_v R\approx 10$ and moderate values of $f_v$ and $R$. For example, it
can be seen in Fig.\ 10 of \cite{fma} that the abundances resulting for
$f_v < 0.2$ strongly deviate from the observed primordial abundances.
Both $f_v R=10$ and $f_v>0.2$ cannot be achieved simultaneously
in the high density scenarios of MFNHS.

Furthermore,
Fig.\ 12 of \cite{fma} shows the limits on the $R$-$\Omega_\mathrm{b}$ parameter
space, due to the observed abundances of $^7$Li and $^2$H. It can be seen
that $\Omega_\mathrm{b}\gtrsim 0.1$ and $R\gtrsim 8$ are ruled out. These
limits do not change significantly when using more modern observational
constraints. Similar conclusions were found in Ref.\ \cite{rau}. Contrary to what
is stated in MFNHS, nucleosynthesis in both proton-rich zones and
neutron-rich zones (created by diffusion) were studied in the latter work.
Although the main idea was to produce heavy elements in the neutron-rich
regions, it turned out that this could not be achieved because of the
limitation on $\eta$ from the light element nucleosynthesis in the high-density,
proton-rich bubbles when compared to observation, even when trying to establish
$\Omega_\mathrm{b}=1$. Similar limitations on $\eta$ should apply for the
calculations of MFNHS.

\begin{table}
\caption{\label{tab:omega}Upper limits of the volume fraction
$f_v^\mathrm{max}$ and lower limits for the density contrast $R^\mathrm{min}$
for several values of $\eta$ and
$\Omega_\mathrm{b}^\mathrm{hi}$, respectively (see text for details).}
\begin{ruledtabular}
\begin{tabular}{rrrr}
\multicolumn{1}{c}{$\eta$}&\multicolumn{1}{c}{$\Omega_\mathrm{b}^\mathrm{hi}$}
&\multicolumn{1}{c}{$f_v^\mathrm{max}$}&\multicolumn{1}{c}{$R^\mathrm{min}$}\\
&&\multicolumn{1}{c}{$(R=\infty)$}&\multicolumn{1}{c}{$(f_v=0)$}\\
\hline
10$^{-6}$ & $83.\dot{3}$ & $6\times 10^{-4}$ & $1.67\times 10^3$\\
10$^{-4}$ & $8333.\dot{3}$ & $6\times 10^{-6}$ & $1.67\times 10^5$\\
10$^{-3}$ & $83333.\dot{3}$ & $6\times 10^{-7}$ & $1.67\times 10^6$
\end{tabular}
\end{ruledtabular}
\end{table}

In order to compare to observational constraints, the nucleosynthesis
products of the high- and low-density regions have to be mixed according to
Eq.\ \ref{eq:mix}. The only indication as to how the light element
abundances relate to the heavy element ones can be found in Fig.\ 7 of
MFNHS. Trying to find a mix reproducing the heavy element abundances on the
level of the ones found in
metal-poor stars or old galaxies will invariably lead to a destruction of
any agreement in the light element abundances because the incompatible
light element production in the high-density regions will dominate the total
abundances $\overline{X}_i$. On the other hand, attempting to find light
element abundances compatible with observational constraints will lead to the
result that the contribution of the high-density bubbles is negligible
($f_v R \ll (1-f_v)$). A more quantitative statement is not possible because
the calculated light element abundances are not quoted in MFNHS.

Finally, it should be noted that there are models of very heavy population
III stars which can account for early re-ionization and abundance patterns
in extremely metal-poor stars without the need of a modified BBN,
e.g.\ \cite{nom1,nom2,nom3}.

Summarizing, while it is still possible that density fluctuations are
introduced into the Early Universe by some mechanism, it has already been
shown in the past that such fluctuations can, if any, only have very limited impact
on BBN. Similar constraints apply to the work of MFNHS.

\end{document}